\documentclass[aps,prd,showpacs,nofootinbib,floats,floatfix,preprintnumbers,groupedaddress,twocolumn]{revtex4-1}
\usepackage{graphicx,epsfig}
\usepackage{dcolumn}
\usepackage{bm}
\usepackage{latexsym}
\usepackage{color}
\usepackage{tabularx}
\usepackage{ulem}
\usepackage{hyperref}
\usepackage{float}
\usepackage{tabularx}
\usepackage{color}
\usepackage{comment}
\usepackage{physics}

\usepackage{tikz}
\usepackage{hyperref}
\usepackage{colortbl}
\usepackage{listings}
\usepackage{amsmath,amsfonts,amssymb}
\usepackage{fancyhdr}
\usepackage{hyperref}
\usepackage{natbib}
\hypersetup{
	colorlinks   = true, 
	urlcolor     = blue, 
	linkcolor    = blue, 
	citecolor    = red 
}

\hypersetup{
	colorlinks   = true, 
	urlcolor     = blue, 
	linkcolor    = blue, 
	citecolor   = red 
}

\begin{document}
\title{Near horizon local instability and quantum thermality}
\author{Surojit Dalui}
\email{suroj176121013@iitg.ac.in}
\author{Bibhas Ranjan Majhi}
\email{bibhas.majhi@iitg.ac.in}
\affiliation{Department of Physics, Indian Institute of Technology Guwahati, Guwahati 781039, Assam, India}

\date{\today}

\begin{abstract}

We revisit our previous proposed conjecture -- {\it horizon creates a local instability which acts as the source of the quantum temperature of black hole}. It is found that a chargeless massless particle moving along the null trajectory in Eddington-Finkelstein (EF) coordinates feels instability in the vicinity of the horizon. Such instability is observer-independent for this particle motion. Moreover, an observer associated with EF coordinates finds the local Hamiltonian as $xp$, where $p$ is the canonical momentum corresponding to the coordinate $x$. Finally, using this Hamiltonian, we notice that at the quantum level, this class of observers feel the horizon as a thermal object with temperature is given by the Hawking expression. We provide this by using various techniques in quantum mechanics and thereby bolstered our earlier claim -- the automatic local instability can be a mechanism for emerging the horizon as a thermal object. In this process, the present analysis provides another set of coordinates (namely EF frame), in addition to our earlier Painleve ones, in which the null trajectory of the massless particle is governed by $xp$ type Hamiltonian in the near horizon regime. 
\end{abstract}

\maketitle

{\section{\label{Intro}Introduction and Motivation}}
The study of thermal and geometrical properties of the black hole horizon and their intimate relationship with the dynamics of particle motion near it is one of the phenomenological interest in recent time. It has been observed that horizon may introduce chaos in a system whenever the system comes under the influence of it \cite{Cardoso:2008bp,Hashimoto:2016dfz,Hashimoto:2018fkb,Dalui:2018qqv,Lu:2018mpr,Cubrovic:2019qee,Dalui:2019umw,Colangelo:2020tpr}. In a recent work \cite{Ma:2019ewq}, a similar thing has been discussed in case of string around charged black brane. In history, chaotic dynamics in the presence of horizon have been extensively studied, but the reason behind this fascinating feature of the horizon is yet to be completely understood. Similarly, why all the horizons (static or stationary) classically give the same phenomenological feature is also an essential question on this note. The investigation has not been limited in the classical scale only, and people have tried to expound the chaotic dynamics of the horizon in the quantum regime as well. The phenomena of quantum chaos are mainly examined by the behaviour of the out-of-time-order correlator (OTOC) of some quantum operator \cite{Maldacena:2015waa,Hashimoto:2017oit}. The characteristic exponential growth of OTOC in those cases is the signature of quantum measure of chaos \cite{Maldacena:2015waa,Hashimoto:2017oit}. 

However, one crucial noticeable point is that, whenever we mention about chaos, there must be some instability factors associated with the system which characterises its chaotic feature. This is known as Lyapunov exponent (see \cite{Strogatz} for a detailed discussion). One recent discovery on the upper bound of the Lyapunov exponent predicted in Sachdev-Ye-Kitaev (SYK) model \cite{Maldacena:2015waa}, has made things very interesting. In the classical picture, it has been found in  \cite{Hashimoto:2016dfz,Dalui:2018qqv} that for any static or stationary black holes, the radial motion of the particle grows exponentially in the near-horizon region. The upper bound of the instability factor, in this case, theoretically comes out to be consistent with that of the SYK model. This has been verified numerically as well {\footnote{There are few cases \cite{Zhao:2018wkl,Guo:2020pgq}, which indeed shows the violation of the bound in Lyapunov exponent. This is either due to considering the unstable equilibrium position of the particle motion far from the horizon \cite{Zhao:2018wkl} or due to the inclusion of the quantum correction in the particle motion, provided by the generalised uncertainty principle \cite{Guo:2020pgq}. Here we shall consider the analysis very near to the horizon. Moreover, in a practical situation, later corrections are very small compared to the original value.}}. In all those cases, the upper bound on the Lyapunov exponent is determined by the surface gravity of the black hole. Since it is well known that the surface gravity is related to Hawking temperature \cite{Hawking:1974rv,Hawking:1974sw}, therefore, this upper bound is dependent on the temperature \cite{Maldacena:2015waa,Shenker:2013pqa}. In a recent paper \cite{Tian:2020bze}, authors have demonstrated that this upper bound of the Lyapunov exponent can be verified in an experimentally realisable setup using a trapped-ion technique.     

In fact, there are shreds of evidence about the connection between the instability of the system and its corresponding quantum thermality. The original work of M.~Srednicki \cite{Srednicki94} suggested that a chaotic system naturally incorporates thermal behaviour. Recently Morita \cite{Morita:2019bfr}, in a similar note, suggested that an unstable classical mode, characterised by a fixed value of Lyapunov exponent, cannot have zero temperature in the quantum scale. One of the extensively investigated unstable systems in this direction is an inverse harmonic oscillator (IHO). At the classical level, the IHO provides instability, and people found that quantum temperature can arise from it, which is determined by the instability factor. A notable feature of this analysis is that the obtained temperature is a pure quantum consequence, and so in the classical limit, it does vanish. All these findings indicate that there is a close connection between the instability and the pure quantum temperature. More precisely, this instability at the classical level can be a source of a pure quantum temperature of a system. 

Contemporary works of Bekenstein \cite{Bekenstein:1973ur} and Hawking \cite{Hawking:1974rv,Hawking:1974sw} indicate that the black hole horizon is a thermodynamic object. Interestingly, the horizon temperature is the observer-dependent quantity and, more importantly, is a pure quantum entity. Since the inception of this thermal concept of horizon, one of the main thrusts till date remains to look for a suitable microscopic origin of aforesaid black hole thermodynamics. There are several attempts, and all of them have their own merits and demerits, and also none of them are complete. Here we want to address one of these important issues in this area. Although the thermodynamical parameter like temperature nicely fits with the horizon, the question remains -- what is the source of this temperature? {\it The underlying mechanism in which sources such temperature still is one of the grey areas}. Motivated by the earlier and recent observations in the context of connection between the classical instability and quantum thermality, we want to explore here such a possibility to explain the existence of horizon temperature. We feel that it can be an important tool to explain this. In this connection, we want to mention that there are some works where IHO (which, as we mentioned earlier,  provides instability) has appeared in the black hole system \cite{Hashimoto:2016dfz,Morita:2019bfr,Hegde:2018xub}. For example, Hashimoto et al \cite{Hashimoto:2016dfz} have shown that if one considers the analysis around the maxima of a field potential in the black hole spacetime, the effective motion of a particle is that in an IHO potential. Later on, Morita \cite{Morita:2019bfr} and Hegde et al \cite{Hegde:2018xub} independently showed that such IHO gives rise to temperature under quantization which is proportional to instability factor of the system. In a completely different context \cite{Maldacena:2005he,Betzios:2016yaq} it has been observed that if a particle scattering phenomenon is considered in a black hole spacetime in the presence of localised shock wave, the effective scattering Hamiltonian comes out to be that of IHO, which also gives rise the same in the quantum regime.

The noticeable fact in all these works, mentioned above, is the possibility of the existence of instability in the form of IHO for a black hole background, which provides thermality to the system. In addition, this feature is {\it local} as it exists in a very small region around a particular point -- either around the maxima of the potential \cite{Hashimoto:2016dfz} or around the location of the shock wave \cite{Betzios:2016yaq}.  Hence none of these analyses is directly connected with the horizon. In other words, {\it the existing observations have not been done around the location of the horizon}. Since we know that the temperature is the property of the horizon, this should arise totally from the investigations around this one-way membrane. That is why our prime objective here is to find if there is any instability near the horizon and try to understand whether that instability is associated with this temperature.

In this regard, we mention that recently such investigation has already been attempted by us \cite{Dalui:2019esx}.  It has been observed that the near horizon Hamiltonian for the radial motion of an outgoing massless particle in the Painleve coordinates is $xp$ type, where $p$ is the conjugate momentum corresponding to coordinate $x$. It takes IHO form in a new canonical conjugate pair of phase space variables and hence is unstable. Moreover, the consequences at the quantum level are found to be the automatic appearance of thermality as long as the Lyapunov exponent of the system remains a positive non-zero quantity. It appeared that the density of states (DOS) is thermal in nature with the temperature is identified as given by the Hawking expression \cite{Hawking:1974rv}. It suggests that the instability, seen by the particle, in the classical scale around the horizon may result in the horizon temperature in the quantum scale. The essential feature of this study is that {\it the system need not be in the chaotic phase; only the unstable feature is enough to get thermality}. In addition, all these are concluded with respect to the {\it Painleve observer}. 

Based on these facts, there are certain remaining issues which are needed to be addressed. They are as follows. 
\begin{itemize}
\item Can the instability be addressed without going to Hamiltonian (or equivalently Lagrangian) analysis?
\item Is there any other set of observers other than Painleve, which also predicts a similar instability and as well as thermality?
\item Are all these features in general observer-dependent, or some are not so?
\item Is it possible to construct Hamiltonian of the system just by the knowledge of the nature of instability in the near horizon regime? This will elaborate on the active role of $xp$ type Hamiltonian in this system.
\item We know that the thermality of the horizon itself is an observer-dependent phenomenon. Can we classify those observers by investigating the connection between instability and thermality? 
\end{itemize}

In this paper, we aim to investigate the whole phenomenon in a more extensive way.  
In the progress of addressing these issues, we find that there is another set of coordinates, namely the Eddington-Finkelstein (EF) coordinates, in which the motion of the particle along the null trajectory also faces the instability in the near horizon regime.  Moreover, such instability is very much there for any observer when the particle is following the outgoing null path in that particular EF coordinates. It implies that {\it the observed near horizon instability of the particle motion is an observer-independent phenomenon for this particular motion of the particle}. Notably,  again the instability factor is given by the surface gravity of the black hole.  Next, we find that the observer associated with the EF frame measures the radial motion as $r\sim e^{\kappa t}$ where $r$ and $t$ are the EF radial and time coordinates, respectively. The corresponding Hamiltonian comes out to be in $xp$ structure. This implies that, as of now,  this particular type of Hamiltonian is observer-dependent -- the frame (either in Painleve or in EF coordinates), which originally defines the particle's motion will see this. 

 Following this classical picture, we next proceed for the quantum calculation. We observe that our present EF observers are suitable to predict the automatic appearance of the thermality as a result of this aforesaid instability. We investigate this fact using different quantum approaches in order to establish our previous conjecture, stated in \cite{Dalui:2019esx}, in a more robust way. In every case, the temperature found out to be that given by the Hawking. Therefore now, under the present investigation,  we reframe this conjecture as --
 \begin{center}
{\it The presence of instability in the near horizon region is the mechanism for providing the temperature to the horizon as seen by a particular class of observers.} 
\end{center}

The paper is organised as follows. In section \ref{Null hypersurface}, we first define the outgoing null path of our massless test particle in EF coordinates. We then analyse the behavior of the trajectory in section \ref{radial} within the classical picture in the near horizon region where the radial trajectories are found to be unstable in nature. Next, using the Raychaudhuri equation \cite{Poisson:2009pwt}, in section \ref{Covariant}, we introduce a technique where we try to realise this instability in a covariant way. Section \ref{Hamiltonian Approach} is devoted to constructing the near horizon Hamiltonian of this particle. Here first, we derive it by using the unstable radial equation of motion of the particle and then verify the same using a direct approach in the context of the dispersion relation. Up to section \ref{Hamiltonian Approach}, every calculation is done in the classical scale. Now, the next section, i.e., Section \ref{QT}, is dedicated to studying the quantum consequences, which is thermality of the horizon in its neighbourhood region. In subsection \ref{Tunneling formalism I}, we start the study of thermality using tunnelling formalism \cite{Srinivasan:1998ty,Parikh:1999mf,Banerjee:2008sn,Banerjee:2009wb,Majhi:2011yi} across the horizon. In the next subsection, we investigate thermality using the detector response approach. Up next in section \ref{BRQNM}, we study the scattering of a massless particle by taking our near horizon Hamiltonian as a pure quantum mechanical scatterer. It again yields thermal nature, and moreover, we are able to extract the imaginary part of the frequency of black hole quasinormal modes (QNM) \cite{Kokkotas:1999bd,Berti:2009kk,Konoplya:2011qq}. In section \ref{Perturbation}, we again study the thermality in the near-horizon region in a perturbative approach considering the obtained near horizon Hamiltonian as a simple quantum mechanical model. In the final section (i.e. Section \ref{Conclusion}), we conclude our work. Three appendices are also included at the end of the paper. In Appendix \ref{App1}, we evaluate the values of the non-affinity parameter, the expansion parameter, and the shear parameter for the null vector in our chosen background. These are essential for our computation of the main work. In Appendix \ref{App3}, we study the detector response function in $(1+1)$ dimensional Schwazchild background for both the outgoing and the ingoing detector, which follow the same path as our test particle. In this case, the near horizon approximation is avoided. In Appendix \ref{Through Gutzwiller}, we re-address the study of thermality through Gutzwiller's formula in order to strengthen our earlier analysis \cite{Dalui:2019esx}. The last two appendices are included mainly for a side discussion.

{\section{\label{Null hypersurface} Outgoing path of massless particle}}
Massless particle follows null-like trajectories and therefore the tangent to the path must be null-like. To identify those, for simplicity, we consider a static spherically symmetric black hole (SSSBH) metric in Schwarzschild coordinates $(t_{s},r,\theta,\phi)$ as
\begin{eqnarray}
ds^2=-f(r)dt_{s}^{2}+\frac{1}{f(r)}dr^{2}+h(r)(d\theta^{2}+\sin^{2}\theta d\phi^{2}).
\label{SSS metric}
\end{eqnarray}
Usually in $(1+3)$ dimensions $h(r)=r^2$, but we kept this as a general function of radial coordinate for our future purpose.
The above coordinate system is singular at the event horizon $\mathcal{H}$, which corresponds to $f(r_{H})=0$. Since we shall confine our investigation in the near horizon regime, the above singularity is not desired to exist in the choice of coordinates. Moreover, we want the particle to follow the outgoing null trajectory. For this purpose, Kruskal-Szekeres (KS) coordinates $(U,V,\theta,\phi)$ in the null-null form will be relevant ones. Since the paths will be outgoing ones, we consider the particle propagates along the normal to $U=$ constant surface, where
\begin{eqnarray}
U=\pm \exp\left(-\kappa u\right)+1~.
\label{K-S coordinates}
\end{eqnarray} 
(Following the discussion in Section 2.5 of \cite{Gourgoulhon:2005ng}, we here choose the above convention).
Since the normal to null surface is tangent to it as well, the particle will propagate along this $U=$ constant surface if the tangent to it's path be this null normal.   
For the above choice, the event horizon $r=r_H$ is located at $U=1$. Here $\kappa$ is the surface gravity defined by $\kappa=f'(r_{H})/2$. The $+(-)$ sign stands for the coordinate is defined outside (inside) the event horizon. For the present purpose, only the $+$ sign will be considered as our test particle resides outside the horizon. In the above, $u$ is known as the Eddington-Finkelstein (EF) outgoing null coordinate. There is also EF ingoing null coordinate $v$. Both of them are related to Schwarzschild coordinates by the relations $u=t_{s}-r_{*}$ and $v=t_{s}+r_{*}$, respectively with the tortoise coordinate $r_{*}$ is defined as
\begin{eqnarray}
dr_{*}=\frac{dr}{f(r)}~.
\label{r_star}
\end{eqnarray}  

The KS coordinates cover the whole spacetime and, therefore, very much adopted to freely falling observer. In order to realize the presence of the horizon by the particle and to confine it outside black hole, the null trajectories will be viewed from a different coordinate system, defined only outside the horizon. For that purpose, we adopt a new set of EF coordinates $(t,r,\theta,\phi)$ where $t$ is related to old coordinates as
\begin{eqnarray}
t=v-r=t_{s}+r_{*}-r~,
\label{E-F timelike}
\end{eqnarray} 
where $r_*$ is taken to be valid outside the horizon. Since $t_s$ and $r$ are timelike and spacelike in the $r>r_H$ region, these new coordinates are properly suited for this region only.
The metric (\ref{SSS metric}), in these, takes the following form:
\begin{eqnarray}
ds^{2}&=&-f(r)dt^{2}+2\Big(1-f(r)\Big)dtdr+\Big(2-f(r)\Big)dr^{2}
\nonumber
\\
&&+h(r)\Big(d\theta^{2}+\sin^{2}\theta d\phi^{2}\Big)~.
\label{E-F metric}
\end{eqnarray}
Considering that the observer is in this frame, we shall calculate all our physical quantity in these coordinates. So now, our next task is to calculate the normal to $U=$ constant surface, which describes the path of the massless particle. Since the observer is in the new EF frame, we need to transform the normal vector in these coordinates. This will give the form of the trajectory of the massless particle with respect to our desire observer.

The normal vector to $U =$ constant (say, $K$) surface is determined by $l_{a}=e^{\rho}\nabla_{a}U$, where $\rho$ is some scalar field on this. For the moment value of $K$ can be {\it any} constant. But since we are interested in near horizon region, at the end, whenever necessary, the limit $U=K\rightarrow 1$ will be taken to achieve our final goal. With this we find the following components of $l^{a}$ on  any $U=$ constant surface in $(t,r,\theta,\phi)$ coordinates as
\begin{eqnarray}
l^{a}=-\kappa e^{\left[\rho-\kappa(t-2r_{*}+r)\right]}\left(1-\frac{2}{f(r)},-1,0,0\right)~.
\end{eqnarray}   
Let us now choose $\rho$ in such a way that $l^{t}=1$. Then we obtain the contravariant components of the tangent to particle trajectory as
\begin{eqnarray}
l^{a}=\left(1,\frac{f(r)}{2-f(r)},0,0\right)~.
\label{lalpha}
\end{eqnarray}
Consequently the covariant components are
\begin{eqnarray}
l_{a}=\left(\frac{f(r)}{f(r)-2},1,0,0\right)~.
\label{l_alpha}
\end{eqnarray}
One can check that on the horizon $\mathcal{H}$ the components reduces to $l^{a} \stackrel{\mathcal{H}}=(1,0,0,0)$, which has the same normalization as that of the timelike Killing vector for this spacetime. This motivated the purpose of above choice for $\rho$.

Now the integral curves $x^{a}(\mu) = (t,r,\theta,\phi)$ of $l^{a}$, characterized by
\begin{eqnarray}
\frac{dx^{a}(\mu)}{d\mu}=l^{a}(x(\mu))~,
\label{tangent of H_U}
\end{eqnarray}
where $\mu$ is the parameter which fixes the particle position at a particular moment, lead to the outgoing null trajectory of our massless particle along any $U=$ constant surface.  Note that the angular components of $l^a$ vanishes and so {\it the particle will have motion only along the radial direction}. In the upcoming section we shall study these trajectories in the near horizon regime, i.e. in the limit $U\rightarrow 1$ (or equivalently $f(r)\rightarrow 0$).

\section{\label{radial} Radial behaviour: instability very near to horizon}
So far, we found the path of our test particle, given by the integral curve (\ref{tangent of H_U}) of the tangent vector (\ref{lalpha}). We are now in a position to investigate the behaviour of this curve in the vicinity of the horizon. Since it has been observed that (\ref{lalpha}) does not have any angular component, the particle will perform only the radial motion.  Therefore, our local analysis will give the nature of the radial coordinate of the particle.

Since the components of tangent vector $l^a$ is given by (\ref{lalpha}) and $x^a=(t,r,\theta,\phi)$, the time component of (\ref{tangent of H_U}) yields
\begin{eqnarray}
\frac{dt}{d\mu}=1\Rightarrow \mu = t~.
\label{E-F time}
\end{eqnarray}
Then the radial component of (\ref{tangent of H_U}) leads to 
\begin{eqnarray}
\frac{dr}{dt}=\frac{f(r)}{2-f(r)}~.
\label{radial component of l}
\end{eqnarray}
The solution of this will give us the behaviour of the particle trajectory in the radial direction.
Since we are interested in the neighbourhood region of the horizon, the metric coefficient $f(r)$ can be taken as the leading term of the Taylor series expansion of it around $r=r_{H}$:
\begin{equation}
f(r)\simeq 2\kappa(r-r_{H})~.
\label{f(r) near horizon}
\end{equation}
Substituting this in (\ref{radial component of l}) and then keeping upto the relevant leading order $(\mathcal{O}(r-r_{H}))$, we obtain
\begin{eqnarray}
\frac{dr}{dt}&\simeq&\frac{2\kappa(r-r_{H})}{2-2\kappa(r-r_{H})}
\nonumber
\\
&\simeq&\kappa(r-r_{H})~.
\label{radial equation}
\end{eqnarray}
The solution of it is 
\begin{equation}
r-r_{H}=\frac{1}{\kappa}e^{\kappa t}~.
\label{radial 1+3D}
\end{equation}
Interestingly, the above analysis indicates the presence of instability in the radial direction of the particle motion as long as the particle is very near to the horizon. Therefore we call this as {\it local} instability. For the rest of the paper, we shall call this as just instability without the explicit mention that it is locally applicable. But keep in mind that whenever such is stated, this is always in a local sense.

Before going into the discussion of the consequences of this local instability, we will show that the above can also be realised in an alternative way. We know that the expansion parameter $\Theta$, defined in Eq. (\ref{define theta}) or in Eq. (\ref{theta}) (see Appendix \ref{App1}), of the null geodesic congruence encodes the information about the behaviour of geodesics -- how the distance between two neighbouring paths changes. Therefore, it is instructive to investigate this parameter in the present context. Below we shall use the value of $\Theta$, calculated in Appendix \ref{App1}, to obtain the behaviour of radial direction in the vicinity of the horizon. We shall come back to this quantity again in the next section.

Use of Eq. (\ref{theta}) of Appendix \ref{App1} for the metric (\ref{E-F metric}) yields
\begin{eqnarray}
\partial_{r}l^{r}= \Theta - \frac{h'(r)}{h(r)}l^{r}+\tilde{\kappa}~,
\label{theta expansion parameter}
\end{eqnarray}  
where  prime indicates the derivative with respective to the $r$ coordinate. Notice that, in the near horizon regime the expression (\ref{theta expression}) for $\Theta$ implies that the expansion parameter is of the order $(r-r_{H})$. Similarly, (\ref{lalpha}) shows $l^{r}$ is also $\mathcal{O}(r-r_{H})$ in this approximation.  On the other hand Eq. (\ref{kappa}) shows that $\tilde\kappa=\kappa+\mathcal{O}(r-r_H)$. Therefore, in the limit $r\rightarrow r_{H}$, keeping only the leading order terms in Eq. (\ref{theta expansion parameter}) we obtain
\begin{eqnarray}
\partial_{r}l^{r}=\kappa~.
\label{partial lr}
\end{eqnarray}
Now using the fact that $l^r=dr/dt$, the solution of the above comes out to be $r=(1/\kappa)e^{\kappa t}+C$ where $C$ is an integration constant. Since, for $r\rightarrow r_{H}$ (i.e. $r_{*}\rightarrow -\infty$) implies $t\rightarrow -\infty$, one obtains $C=r_{H}$. Therefore we have the same solution (\ref{radial 1+3D}).

We now make a comment for the same in $(1+1)$ dimension static black hole case. This is needed as later in some situations; we shall consider this lower-dimensional case for the simplicity of the calculation (For example, a side discussion has been made in Appendix \ref{App3} in this spacetime dimensions).  For (1+1) dimensional case i.e. considering the $(t-r)$ sector of metric (\ref{E-F metric}), one can readily show that time and radial components of $l^a$ are given by those given in (\ref{lalpha}). Therefore, one again finds the same radial behaviour as obtained in (\ref{radial 1+3D}). Also, as here $\Theta$ vanishes and the determinant of the metric is $g=-1$, the definition for expansion parameter (\ref{theta}) reduces to Eq. (\ref{partial lr}). Hence one finds (\ref{radial 1+3D}) again, and so the existence of the instability in the particle motion in the near horizon region persists in this case as well. This indicates that the present instability is completely due to the influence of the horizon in spacetime, not specific to the number of spacetime dimensions.

{\section{\label{Covariant} A covariant realisation of local instability}}
We found that the radial motion is unstable in nature in the very near to the horizon. In this regard, it is natural to ask -- what happens to the family of these null geodesics in this region? Whether this congruence of geodesics also faces a similar instability due to the horizon. Moreover, in the last section we mentioned that the expansion parameter $\Theta$ can be an important quantity to illuminate our main investigation. Particularly as it measures the separation between the two nearby geodesics, it will be interesting to see how this separation changes with time. Thus we shall have a more concrete idea of instability, provided by the horizon. Therefore, the present section will be dedicated to examining the evaluation of $\Theta$ for null geodesics in the nearby region of the horizon. The most promising way is to start with the Raychaudhuri's equation for null congruence \cite{Poisson:2009pwt}. Since it is in the covariant form, we expect that the evaluation character of $\Theta$, obtained from this, in contrary to the earlier section, may provide a covariant description of our aforesaid instability. 

 Raychaudhuri equation for null geodesics is \cite{Poisson:2009pwt}
\begin{equation}
\frac{d\Theta}{d\mu}=\tilde{\kappa}\Theta-\frac{1}{2}\Theta^2 - \sigma_{ab}\sigma^{ab}+\omega_{ab}\omega^{ab}-R_{ab}l^al^b~.
\label{Raychaudhuri eqn}
\end{equation}
Here we shall study this equation in the near horizon of our SSSBH spacetime (\ref{E-F metric}). All the quantities are defined with respect to the null vector (\ref{lalpha}). Let us now examine each of the terms on the right hand side of the above equation. These are all calculated in Appendix \ref{App1}. We found that the shear parameter $\sigma_{ab}=0$ (see Eq. (\ref{sigma})) and since $l_{a}$ is hypersurface orthonormal, we must have the rotation parameter $\omega_{ab}=0$ as well. Next note that in the near horizon region $\Theta\sim \mathcal{O}(r-r_{H})$ (see Eq. (\ref{theta expression})), whereas as mentioned in the last section, $\tilde\kappa=\kappa+\mathcal{O}(r-r_H)$.  The evaluation of the term $R_{ab}l^{a}l^{b}$ for metric (\ref{E-F metric}) yields
\begin{eqnarray}
R_{ab}l^{a}l^{b}=\frac{f^2(r)\left(h'^{2}(r)-2h(r)h''(r)\right)}{2(f(r)-2)^{2}h^2(r)}~.
\label{Rab la lb}
\end{eqnarray}
Now, for the value of $h(r)=r^{2}$ the above term vanishes. Therefore, keeping the leading order terms, i.e. $\mathcal{O}(r-r_H)$ terms in the right hand side of Eq. (\ref{Raychaudhuri eqn}) one obtains
\begin{eqnarray}
\frac{d\Theta}{d\mu}=\kappa\Theta\label{Raychaudhuri eqn near horizon}~.
\end{eqnarray}  
Performing the integration of the above equation we obtain the form of the expansion parameter as 
\begin{eqnarray}
\Theta=\kappa e^{\kappa \mu}
\label{theta from Raychaudhuri eqn}~.
\end{eqnarray}
This implies that in the near horizon the expansion of geodesic congruences is exponentially increasing as $\mu$ increases. It characterises the presence of instability in the geodesic motion of the particle. 

This analysis not only indicates the presence of local instability for the particle motion but also provides a covariant description and realisation of this phenomenon. Since $\Theta$ is a scalar quantity, we now understand that for this particular particle motion, the aforesaid {\it instability is an observer independent feature of horizon}.

Just for completeness, we now show that from Eq. (\ref{theta from Raychaudhuri eqn}) the explicit form of unstable nature in radial motion can be evaluated. In our EF coordinates we identified $\mu=t$ (see Eq. (\ref{E-F time})). One can check that at the horizon (i.e. $t\rightarrow -\infty$), $\Theta$ vanishes, which implies that the above solution correctly satisfies the required boundary condition.
The value of $\Theta$ for our metric (\ref{E-F metric}) is given by (\ref{theta expression}). In the near horizon regime, at the leading order, it comes out to be
\begin{equation}
\Theta \simeq \frac{2\kappa}{r_H}(r-r_H)~,
\label{B1}
\end{equation}
where we have used $h(r)=r^2\rightarrow r_H^2$ and $h'(r)=2r\rightarrow 2r_H$. Substitution of this in the solution (\ref{theta from Raychaudhuri eqn}) yields $r-r_H\simeq(r_H/2)e^{\kappa t}$. Thus again, we found the similar unstable nature in the radial direction. It must be mentioned that {\it although the instability is an observer-independent feature, this particular radial character with time is related to EF observer}. This is a very crucial observation in this analysis. We shall talk about more on this in our later discussion. It plays a big role in the concept of observer-dependent thermality, which will be introduced in the upcoming sections.

{\section{\label{Hamiltonian Approach}{Near horizon instability: Hamiltonian analysis}}
So far, without using any formal prescription, like Lagrangian or Hamiltonian analysis, we have been able to show the appearance of local instability on the radial motion of a massless particle in the vicinity of the horizon. This feature has been shown earlier in \cite{Dalui:2019esx} using the Hamiltonian analysis using Painleve coordinates for the metric. It was shown that the near horizon Hamiltonian takes the form $\sim xp$, where $x=r-r_H$ and $p$ is the radial momentum.  In this work, we are using a new EF coordinates and found that here also, a similar feature is appearing in radial motion even in this new coordinates. Therefore, it would be interesting to see whether a Hamiltonian prescription can be built out in our present analysis. More importantly, we are interested in investigating the possibility of finding out the Hamiltonian of our system using the obtained radial feature in the earlier sections. If so, then whether it is again similar to $xp$. In this section, we shall first find the Hamiltonian from our earlier findings on the radial trajectory and then verify this by deriving the same using dispersion relation for the massless particle on the background (\ref{E-F metric}). This obtained structure of Hamiltonian will be very important for the later purpose of our analysis.

\subsection{Hamiltonian from trajectories}
The near horizon radial motion is driven by Eq. (\ref{radial 1+3D}). Therefore use of Hamilton's equation of motion $\dot{x}=\partial H/\partial p$ implies
\begin{equation}
\frac{\partial H}{\partial p} = \kappa x~,
\label{B2}
\end{equation}
where $x \equiv r-r_H$.
Solution of this is given by $H=\kappa x p + f_1(x)$, where $f_1(x)$ is an arbitrary function of radial coordinate. This can be fixed by using the information that the corresponding Lagrangian must vanish as we are dealing with massless free particle. The Lagrangian for this solution comes out to be
\begin{equation}
L= p\dot x -H = -f_1(x)~.
\label{B3}
\end{equation}
So to make it vanish, we must choose $f_1(x)=0$. Thus we find that the Hamiltonian in the near horizon regime is given by
\begin{eqnarray}
H = \kappa xp~.
\label{xp Hamiltonian}
\end{eqnarray}
We now verify this below by direct evaluation of Hamiltonian from the dispersion relation. This method has been adopted earlier in \cite{Dalui:2019esx}, but for Painleve coordinates.

\subsection{Hamiltonian from dispersion relation}
We again start with the static spherically symmetric metric written in EF coordinates (\ref{E-F metric}) which has a timelike Killing vector $\chi'^{a}=(1,0,0,0)$ and the energy of a particle moving under this background is given by $E=-\chi'^{a}p_{a}=-p_{t}$, where $p_{a}$ is the four momentum whose components are $p_{a}=(p_{t},p_{r},0,0)$. The angular components are chosen to be zero as for our choice of path there is only radial motion (see Eq. (\ref{lalpha})). Using the covariant form of the dispersion relation $g^{ab}p_{a}p_{b}=0$ for massless particle, we obtain the equation of the energy in terms of the radial component of the momentum as
\begin{eqnarray}
(f(r)-2)E^{2}-2(1-f(r))Ep_{r}+f(r)p_{r}^{2}=0~.
\label{dispersion relation}
\end{eqnarray}
It is found that the energy has two solutions:
\begin{eqnarray}
E=\frac{(f(r)-1)p_{r}\mp p_{r}}{2-f(r)}~,
\label{energy eqn}
\end{eqnarray}
where the positive sign for the outgoing particle and the negative sign for the ingoing one. With the near horizon approximation i.e. for $f(r)$ given in Eq. (\ref{f(r) near horizon}), we obtain the expression for the energy of the outgoing particle (i.e. taking the $+ve$ sign solution) as
\begin{eqnarray}
E&=&\frac{(f(r)-1)p_{r}+p_{r}}{2-f(r)}
\nonumber
\\
&=&\frac{\kappa(r-r_{H})p_{r}}{1-\kappa(r-r_{H})}
\nonumber
\\
&\simeq &\kappa(r-r_{H})p_{r}+\mathcal{O}(r-r_{H})^{2}~.
\label{E outgoing}
\end{eqnarray}
Since we are interested near to the horizon, taking up to the first-order one obtains the expression of the Hamiltonian for the outgoing particle as (\ref{xp Hamiltonian}) {\footnote{It may be mentioned that this type of Hamiltonian is somehow very common feature of the gravitational system. It appears in different situations in the presence of gravity. At the thermodynamic level, the surface part of the Einstein-Hilbert action yields $xp$ type Hamiltonian \cite{Bakshi:2016vpp}. Also similar observation has been noticed for the dynamics of super-translational parameter in the context of asymptotic symmetry of a null surface \cite{Maitra:2019eix}.}}, with $p_{r}\equiv p$. 

So we observed that the nature of Hamiltonian, like in Painleve coordinates, is $\sim xp$ even in EF coordinates. 
This is inherently unstable in nature, having the hyperbolic points at $x=0$ and $p=0$, which induces the instability into the particle's motion. The solutions of the equations of motion corresponding the Hamiltonian (\ref{xp Hamiltonian}) are
\begin{eqnarray}
x(t)=x(0)e^{\kappa \lambda};~~~~~~p(t)=p(0)e^{-\kappa \lambda}~,
\end{eqnarray}
where $\lambda$ is the affine parameter, which defines the momentum of the particle as $p^r=dr/d\lambda$.
It immediately shows us again that at the classical level, the radial motion of the massless particle is unstable in the vicinity of the horizon, which we have already shown in different approaches in the previous sections.

We shall end this classical discussion with the following comment.
Through various approaches, people have already seen that horizon may induce chaos in a system whenever the system comes under the influence of it \cite{Hashimoto:2016dfz,Dalui:2018qqv,Dalui:2019umw} and notably this is common to any black hole spacetime. To follow up the real cause of this universal feature, we argued in \cite{Dalui:2019esx} that the instability is the main cause of it. There we studied the particle motion in the Painleve coordinates.  Here we showed that the same instability appears in EF coordinates. Moreover, such is observer-independent, provided that the particle is following a particular null path in the near-horizon region. 
But it must be remembered that the particular radial nature of the particle trajectory is observer-dependent. Since the behaviour of a system under the influence of the horizon has been studied with these trajectories, it may happen that the appearance of chaos is observer-dependent phenomenon. This last statement is not conclusive at this stage; rather, it is a suggestive one. We need more investigation in this direction to reach any definite conclusion. 


\section{\label{QT}Quantum thermality}
Till now, we observed that at the classical level, the horizon creates a local instability on the radial motion of a massless particle. This is completely a local phenomenon as it may not be observed when the particle motion is considered over the full spacetime. We are now curious to know whether such a local phenomenon can have any observable consequence. In this section quantum aspects will be addressed. Two of the authors of this paper already showed in \cite{Dalui:2019esx} that the instability may provide temperature to the horizon with respect to the Painleve observer. Here we observed that our EF frame also perceives similar instability in the trajectory. Therefore we will again investigate if this can again explain the thermality of the horizon. Here our main objective is to study the consequences of aforesaid instability at the quantum level in various possible ways in order to verify the robustness of the aforesaid thermality. It will be found that quantum thermality of the horizon is unavoidable and thereby providing a robust evidence of our earlier claim.

\subsection{\label{Tunneling formalism I} Tunneling formalism}
Classically nothing can escape from the black hole. But the quantum probability of escaping from the barrier of the horizon can be different. The previously obtained Hamiltonian can be used here to find this. It is the main quantity which is found in tunneling formalism to study the Hawking effect (For the underlying concept and details of this method see \cite{Srinivasan:1998ty,Parikh:1999mf,Banerjee:2008sn,Banerjee:2009wb}. Also see \cite{Majhi:2011yi} for an extensive list of works on tunneling formalism) {\footnote{Ref. \cite{Srinivasan:1998ty} adopted the Hamilton-Jacobi method whereas Ref. \cite{Parikh:1999mf} based on null geodesic approach. Based on the tunneling idea, using the connections between the coordinates on both sides of the horizon, the same has been done in \cite{Banerjee:2008sn,Banerjee:2009wb}.}}. Adopting the concept of this mechanism, here we shall calculate the tunneling probability of a particle. The analysis is semi-classical in nature, and calculation at the vicinity of the horizon is sufficient.

We start with the standard ansatz for wave function for a particle as
\begin{eqnarray}
\Psi(x)=\exp\left[\frac{i}{\hbar} S(x)\right]~,
\label{phi}
\end{eqnarray}
where $S(x)$ is the Hamilton-Jacobi action for the particle, defined as an integration of the momentum $p$ of the particle with respect to the position coordinate $x$ variable:
\begin{equation}
S(x)=\int pdx~.
\label{QB1}
\end{equation}  
(Here we have considered the above expression for two-dimensional phase space). The outgoing and ingoing trajectories correspond to $\partial S/\partial x>0$ and  $\partial S/\partial x<0$, respectively. 
For our present situation, both the outgoing and ingoing particles are just outside the horizon. Therefore we are interested in calculating the absorption probability of the outgoing particle while the emission probability for ingoing one. The ratio of them will give us the required tunneling probability.

The energy of the outgoing particle is given by (\ref{xp Hamiltonian}). Since $H=E$ is the conserved quantity here, we substitute $p$ in terms of $x$ in (\ref{QB1}) to find the outgoing action. Also since the absorption  probability will be our main interest, the limits of the integration must be chosen $x=\epsilon$ to $x=-\epsilon$ where $\epsilon>0$ (i.e. from just outside the horizon to just inside). Thus the ``absorption'' action is given by
\begin{eqnarray}
S [\textrm{Absorption}]&=&\frac{E}{\kappa}\int_{\epsilon}^{-\epsilon}\frac{dx}{x}
\nonumber
\\
&=&-\frac{i\pi E}{\kappa} + (\text{real part})~.
\label{S out}
\end{eqnarray} 
In performing the above integration, we noticed that $x=0$ is the pole of the integrand. To evaluate it, the lower complex plane is being considered. Observe that since the particle starts from outside the black hole where $x>0$, we have $\partial S/\partial x>0$, which is consistent with the definition of the outgoing nature of the trajectory. On the other hand, the ``emission'' action for the ingoing particle will be real as the limits of integration never include the horizon singularity. This can be checked trivially with the identification of energy for the ingoing particle as $E=-p$ (see Eq. (\ref{energy eqn})).
So, the probability of absorption turns out to be 
\begin{eqnarray}
P[\textrm{Absorption}]&\sim& \Big|e^{\frac{i}{\hbar}S[\textrm{Absorption}]}\Big|^2 
\nonumber
\\
&\varpropto& \exp\left(\frac{2\pi E}{\hbar\kappa}\right)~.
\label{Pout}
\end{eqnarray} 
whereas the probability of emission is $P[\textrm{Emission}]=1$.
Hence the tunneling probability is evaluated as
\begin{eqnarray}
\Gamma=\frac{P[\textrm{Emission}]}{P[\textrm{Absorption}]}\sim \exp\left(-\frac{2\pi E}{\hbar\kappa}\right)~.
\label{Pout/Pin}
\end{eqnarray}
Note that the above one is thermal in nature. 
The temperature is identified as
\begin{eqnarray}
T=\frac{\hbar\kappa}{2\pi}~.
\label{TH}
\end{eqnarray}
This temperature exactly matches with the standard Hawking expression \cite{Hawking:1974rv} for black hole. 

We just observed that the near horizon Hamiltonian (\ref{xp Hamiltonian}) predicts a finite probability of escaping a particle from the horizon and thereby providing a temperature to the horizon. Since this Hamiltonian shows a local instability, we argue that such instability is responsible for the thermal behaviour of the black hole. From this analysis, we can note that the observer is associated with the EF coordinates. But at this point, it is not vivid whose vacuum state is filled with a particle with respect to this frame. It is the well-known limitation of the tunneling approach. This will be illuminated in the next subsection by adopting a different approach.
 
\subsection{\label{Near horizon detector response}Detector's response}
Thermality is an observer dependent phenomenon \cite{Unruh:1976db,Book1} and vacuum plays an important role in this case. The precise choice of observer and the corresponding choice of vacuum is very important in that sense. Therefore, the aim of our next approach is to identify the observer and the corresponding vacuum state connected to this thermality. One such popular approach is investigating through the detector's response of a two-level atomic detector, which can give us the clear idea to identify our observer and the vacuum. The choice of the observer here is the one which is following the path (\ref{radial 1+3D}) in the near horizon regime. The vacuum is chosen to be Boulware vacua, which is defined with respect to the static observer in Schwarzschild coordinates. We will find the transition rate of the atomic detector, which moves along the trajectory (\ref{radial 1+3D}) with respect to this Boulware vacuum. The calculation must be performed very near to the horizon. The particular preference of this vacuum among others like Unruh or Kruskal vacua is due to the fact that Unruh and Kruskal ones are not vacuum with respect to static frame, whereas Boulware is a trivial one. Therefore it is apparent that the present moving frame again finds Unruh and Kruskal vacua as non-trivial one. Hence whether Boulware appears to be non-trivial with respect to our present observer will be an interesting observation (a discussion of defining different vacuum states can be followed from \cite{Book1}).

Let us consider a two-level atomic detector (say $a$ is the excited level and $b$ is the ground state) is moving along the geodesic (\ref{radial 1+3D}).  We consider the massless scalar field $\Phi$ under this background and its modes are denoted by $u_{\nu}$ with frequency $\nu$. The modes for the atomic detector we denote as $\psi_{\omega}$, where $\omega$ is the characteristic frequency. The interaction Hamiltonian between the atomic detector and the field is taken as
\begin{eqnarray}
\hat{H}_{int}(\tau)=Q[(\hat{a}_{\nu}u_{\nu}+h.c.)(\hat{\sigma}_{\omega}\psi_{\omega}+h.c.)]~,
\label{interaction hamiltonian}
\end{eqnarray}
where the operator $\hat{a}_{\nu}$ is the photon annihilator operator and $\hat{\sigma}_{\omega}$ is the atomic detector lowering operator. $h.c.$ signifies the hermitian conjugate. $Q$ is the coupling constant which determines the strength of the interaction and $\tau$ is the detector's clock time. This type of model was originally considered for this purpose in \cite{Scully:2017utk} and later has been subsequently used in \cite{Chakraborty:2019ltu}.

Initially, when there is no photon is detected, the detector stays in the ground state $|b\rangle$ i.e the field is in the Boulware vacuum $|0\rangle$. So the initial state of the whole system is $|0,b\rangle = |0\rangle\otimes |b\rangle$. Now after interaction the detector will go to state $|a\rangle$. Then the transition amplitude of the detector, using the first order perturbation theory, is given by
\begin{eqnarray}
\Gamma= -i\int_{\tau_{i}}^{\tau_{f}}d\tau\langle 1_{\nu},a|\hat{H}_{int}(\tau)|0,b\rangle~,
\label{amplitude of detector}
\end{eqnarray} 
where  $|1_{\nu}\rangle$ is the one particle state of $\Phi$.  In this subsection, we have chosen $\hbar=1$.
In this case the probability of the excitation of the atomic detector for interaction Hamiltonian (\ref{interaction hamiltonian}) becomes (see \cite{Chakraborty:2019ltu} for detailed calculation)
\begin{eqnarray}
P_{\uparrow}=Q^{2}\bigl\lvert\int_{r_{i}}^{r_{f}}dr\left(\frac{d\tau}{dr} \right)u_{\nu}^{*}(r)\psi_{\omega}^{*}(r)\bigr\rvert^{2}~.\label{probability of detection in radial form}
\end{eqnarray}
Here re-expressing the detector's path (\ref{radial 1+3D}) as $t$ in terms of radial coordinate we obtain      
\begin{eqnarray}
t=\frac{1}{\kappa}\ln\left(\frac{r}{r_{H}}-1\right) + \text{constant}~,
\label{time expr for the detector}
\end{eqnarray}
where the constant, irrelevant for the present analysis, is given by $(1/\kappa)\ln(\kappa r_H)$.  
Next taking $\tau=t$ the positive frequency mode corresponding to the detector is 
\begin{eqnarray}
\psi_{\omega}=e^{-i\omega t}~.
\label{detector mode}
\end{eqnarray}
The positive frequency Boulware mode for massless scalar field can be obtained by solving the Klein-Gordon (KG) equation $\square\Phi=0$ under the background of (\ref{SSS metric}). Near the horizon KG equation reduces to 
\begin{eqnarray}
\left[\frac{\partial^{2}}{\partial t^{2}_{s}}-\frac{\partial^{2}}{\partial r^{2}_{*}}\right]\Phi = 0~,
\label{KG eqn}
\end{eqnarray} 
where in the near horizon limit $r_*$ is given by
\begin{eqnarray}
r_{*}\simeq\frac{1}{2\kappa}\ln{\left(\frac{r}{r_{H}}-1\right)}~.
\label{r_star for schild}
\end{eqnarray} 
The solutions are $e^{-i\nu(t_{s}\pm r_{*})}$, where the positive sign corresponds to ingoing and the negative sign refers to outgoing modes. Here, our detector is moving in the outward direction and so we will consider the ingoing Boulware mode to investigate the response of the detector. Therefore we choose
\begin{eqnarray}
u_{\nu}=e^{-i\nu(t_{s}+r_{*})}~.
\label{mode solution for photon}
\end{eqnarray}   
Hence expressing the integrand of Eq. (\ref{probability of detection in radial form}) in terms of the radial coordinate and using (\ref{time expr for the detector}), we obtain the probability of transition as 
\begin{eqnarray}
P_{\uparrow}=\frac{Q^{2}}{\kappa^{2}}\Big| \int_{r_{H}}^{r_{f}}d\left(\frac{r}{r_{H}}\right)\left(\frac{r}{r_{H}}-1\right)^{\frac{i}{\kappa}(\nu + \omega)-1} e^{i\nu r}\Big|^{2},
\label{probability of detection by outgoing detc}
\end{eqnarray}
where the upper limit is taken as position $r_{f}$ which is situated very near to the horizon. It has to be chosen in such a way that it's value satisfies our near horizon approximation i.e. $\left(\frac{r}{r_{H}}-1\right)<<1$.

In order to get some convenient look of Eq. (\ref{probability of detection by outgoing detc}) let us first make change of variable: $(r/r_{H})-1=y$. Then (\ref{probability of detection by outgoing detc}) reduces to 
\begin{eqnarray}
P_{\uparrow}=\frac{Q^{2}}{\kappa^{2}}\Big|\int_{0}^{y_{f}}dy~y^{\frac{i}{\kappa}(\omega+\nu)-1}e^{i\nu (y+1)}\Big|^{2}~.
\label{probability of detection y form}
\end{eqnarray} 
This can be expressed in terms of lower incomplete Gamma function (See page no. 527 of ref. \cite{Arfkenbook})
\begin{eqnarray}
P_{\uparrow}=\frac{Q^{2}}{\kappa^{2}}\Bigg|\frac{1}{(-i\nu)^{\frac{i}{\kappa}(\omega+\nu)}}~\gamma\Bigg(\frac{i}{\kappa}(\omega+\nu),-i\nu y_{f}\Bigg)\Bigg|^{2}~.
\end{eqnarray}
But to get a better understanding, here we shall examine it numerically for different values of $\omega$. In order to do that, first we need to make all the variables dimensionless. We choose the following substitutions in Eq. (\ref{probability of detection y form}):
\begin{eqnarray}
r_{H}\omega=\omega';~~~~r_{H}\nu=\nu'~~~~\text{and}~~~~r_{H}\kappa=\kappa'.
\end{eqnarray}      
Then Eq. (\ref{probability of detection y form}) reduces to the following form:
\begin{eqnarray}
P'_{\uparrow}=\Big|\int_{0}^{y_{f}}dy~y^{\frac{i}{\kappa'}(\omega'+\nu')+\epsilon -1}e^{i(\nu'+ i\epsilon) (y+1)}\Big|^{2}~.\label{probability of detection with epsilon}
\end{eqnarray} 
where $P'_{\uparrow} = \frac{\kappa'^{2}}{Q^{2}r_{H}^{2}}$ and in the above we have introduced a very small parameter $\epsilon$ to make the integration convergent. 

Now, we numerically integrate the above expression for different values of $\omega'$ and then plot $\nu'^{2}P'_{\uparrow}$ as a function of $\nu'$. The plot is represented in Fig. (\ref{Fig1}). 
\begin{figure}[h!]
\begin{centering}
\includegraphics[scale=0.4]{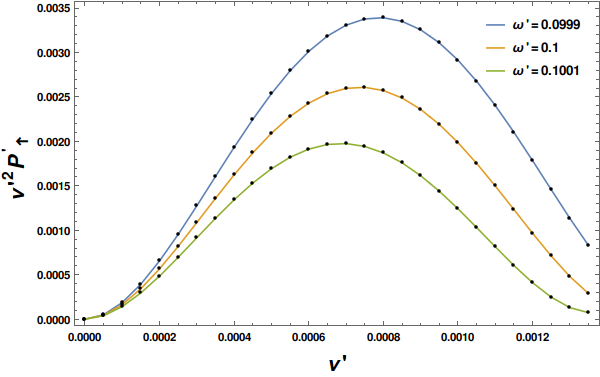}
\end{centering}
\caption{Plot of $\nu'^{2}P'_{\uparrow}$ Vs $\nu'$ for different values of $\omega'$. The choice of the small parameter is $\epsilon=0.00095$.}
\label{Fig1}
\end{figure}
This shows that the nature of the transition probability of the detector is similar to Planck distribution. So the detector will register particles in the Boulware vacuum when it moves along our local unstable path.  Hence with respect to this observer, the vacuum appears to be thermal. As we increase the value of $\omega'$, the peak of the curve decreases. It means that for higher values of $\omega'$, the probability of detecting particle gets lessened.

In the similar approach one can also derive the expression for probability of detecting an outgoing scalar field mode by an ingoing detector in the near horizon region. It means that the detector is moving very near to the horizon but this time its direction of motion is towards the horizon, just opposite to the previous case. In this case the EF time coordinate is represented in terms of the outgoing EF coordinates $(u,r,\theta,\phi)$ as
\begin{eqnarray}
t=u+r=t_{s}-r_{*}+r\label{E-F time outgoing}~.
\end{eqnarray}
In the similar approach we can re-express the path of the ingoing detector as $t$ as a function of the radial coordinate. For the near horizon approximation using Eq. (\ref{lalpha out}) of the Appendix \ref{Detector response out} we obtain
\begin{eqnarray}
t=-\frac{1}{\kappa}\ln{\left(\frac{r}{r_{H}}-1\right)}+\text{constant}~.\label{t ingoing detector near horizon}
\end{eqnarray}   
Now, proceeding with exactly similar approach like the case of the outgoing detector one can land up to the expression of probability which turns out to be
\begin{eqnarray}
P_{\downarrow}=\frac{Q^{2}}{\kappa^{2}}\Big|\int_{0}^{y_{f}}dy~y^{-\frac{i}{\kappa}(\omega+\nu)-1}e^{-i\nu (y+1)}\Big|^{2}\label{probability of detection ingoing detcr y form}~,
\end{eqnarray}   
and it basically gives the same result as in the case of the outgoing detector (FIG.(\ref{Fig1})).

Therefore, the outgoing and the ingoing atomic detector, following the null path, detects ingoing and outgoing scalar particle, respectively, in the Boulware vacuum. The Planckian nature of the plots suggests that at the quantum level, the vacuum appears to be thermal. We showed this for near horizon trajectory. For completeness, we also show that our present observer, when moves throughout the whole spacetime, then also it will perceive thermality. This we present in Appendix \ref{App3} for the Schwarzschild black hole where the near horizon approximation is being avoided.  To have a complete analytic analysis, the calculation is performed in $(1+1)$ dimensions, and the temperature is identified to be the Hawking expression.

\section{\label{BRQNM}Scattering of particle and QNM}
Till now, we observed that the horizon provides an unstable potential to the massless particle in its neighbourhood region. Moreover, it causes the particle to feel the black hole as a thermal object. This quantum phenomenon can also be elaborated through a ``scattering'' model of a particle. The idea is the following. When a particle is moving very near to the horizon, it will feel the influence of the horizon through the local Hamiltonian (\ref{xp Hamiltonian}). Then the state of the particle will be influenced. The change of wave function can be evaluated by visualising  (\ref{xp Hamiltonian}) as the governing potential for the scattering phenomenon. In order to proceed towards the main purpose first, we need to identify the initial (before scattering) and final (after scattering) energy eigenstates of the system. 

Our $xp$ Hamiltonian can be visualised as that for an inverted harmonic oscillator (IHO) in a new set of canonical variables $(X,P)$. The relation between the old and these new ones are $x=\frac{1}{\sqrt{2}}(P-X)$ and $p=\frac{1}{\sqrt{2}}(P+X)$. Then in the $(X,P)$ diagram, the old $(x,p)$ variables are considered to be as ingoing and outgoing coordinates, respectively. This is shown in Fig. \ref{Fig2}.
\begin{figure}[h]
\begin{center}
\includegraphics[scale=0.2]{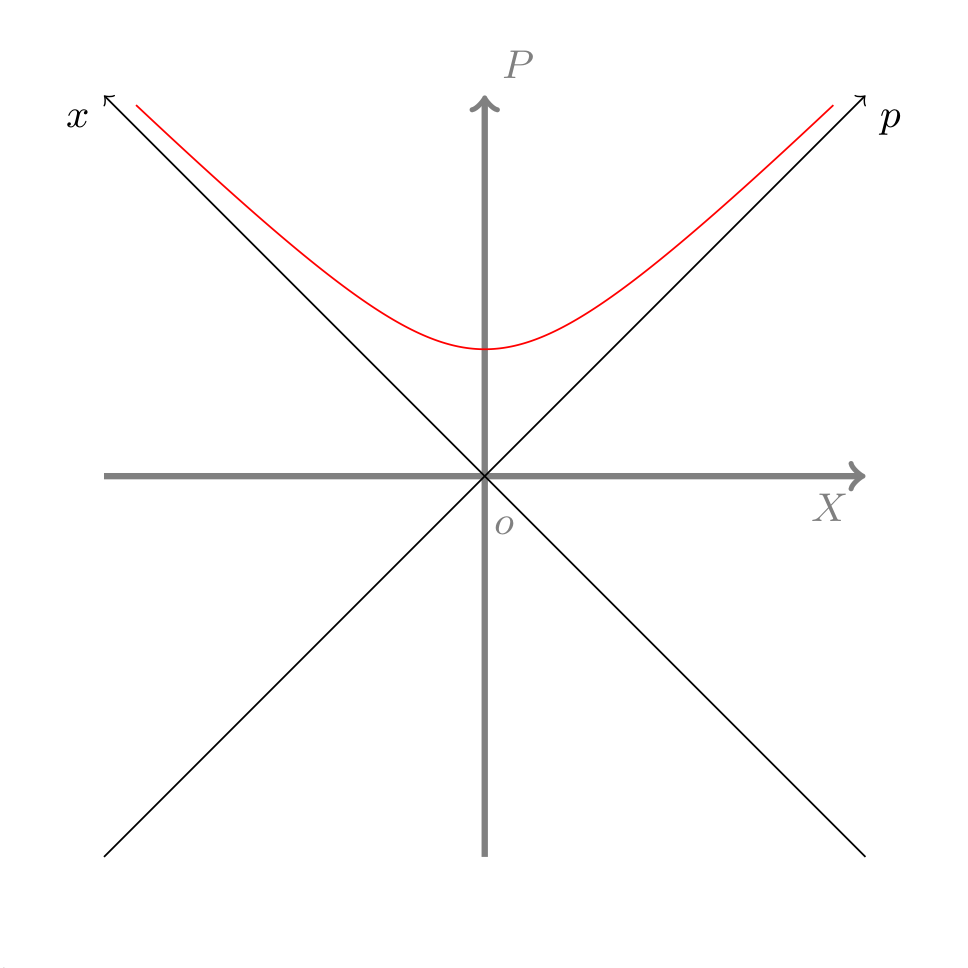} 
\end{center}
\caption{$X-P$ diagram: red line represents the trajectory of the particle.}
\label{Fig2}
\end{figure}
Here the value of $x$ is always positive, therefore the trajectory for $E>0$ in the $X-P$ plane always remains in that quadrant where both $p$ and $x$ are positive definite (see Fig. (\ref{Fig2})). Since $x$ is identified as the ingoing coordinate, the energy eigenstate in $x$ representation is the initial state of the system. Likewise, $p$ representation energy state is our final state for the system. Therefore our next task is to find the eigenstates for Hamiltonian (\ref{xp Hamiltonian}) in both representations.

In order to make Hamiltonian (\ref{xp Hamiltonian}) hermitian, we express this as 
\begin{eqnarray}
\hat{H}=\frac{\kappa}{2}\left[\hat{x}\hat{p}+\hat{p}\hat{x}\right]~,
\label{xp amiltonian}
\end{eqnarray}
where the basic commutator is given by $[\hat{x},\hat{p}]=i\hbar$.
Then the inner product between $x$ and $p$ states is
\begin{eqnarray}
\langle x|p\rangle = \frac{1}{\sqrt{2\pi\hbar}}\exp\left(\frac{ixp}{\hbar}\right)~.
\label{xp inner product}
\end{eqnarray}
To find the initial state, we represent  Hamiltonian operator in position representation: 
\begin{eqnarray}
\hat{H}=-i\tilde{\gamma}\left[x\frac{\partial}{\partial x}+\frac{1}{2}\right]~,
\label{xp in coordinate rep}
\end{eqnarray}
where $\tilde{\gamma}=\hbar\kappa$. With this the initial state with energy $E$ comes out to be
\begin{eqnarray}
\langle x|E\rangle_{i}=\frac{1}{\sqrt{2\pi\tilde{\gamma}}}\frac{1}{x^{\frac{1}{2}-\frac{iE}{\tilde{\gamma}}}}~~~ {\textrm{with}} ~~~x>0~.
\label{energy eigenstate x rep}
\end{eqnarray} 
The final state is determined by expressing the Hamiltonian in momentum representation:
\begin{eqnarray}
\hat{H}=i\tilde{\gamma}\left[p\frac{\partial}{\partial p}+\frac{1}{2}\right]~.
\label{xp in p rep}
\end{eqnarray}
Eigenstate with energy $E$ of this operator yields the final state of the system as 
\begin{eqnarray}
\langle p|E\rangle_{f} = \frac{1}{\sqrt{2\pi\tilde{\gamma}}}\frac{1}{p^{\frac{1}{2}+\frac{iE}{\tilde{\gamma}}}} ~~~ {\textrm{with}} ~~~p>0~.
\label{energy eigenstate p rep}
\end{eqnarray}

Now we shall find the relation between the final state and initial state.  This is done as follows:
\begin{eqnarray}
\langle x|E\rangle_{f}&=&\int_{-\infty}^{\infty}dp\langle x|p\rangle\langle p|E\rangle_{f}
\nonumber
\\
&=&\frac{1}{2\pi\sqrt{\tilde{\gamma}\hbar}}\int_{0}^{\infty} dp~~ e^{\left(\frac{ix}{\hbar}\right)p}~p^{\left(-\frac{iE}{\tilde{\gamma}}+\frac{1}{2}\right)-1}~.
\label{BR2}
\end{eqnarray} 
To perform the integration, we use the formula (see page no. 604 of \cite{Paddybook} for details)
\begin{equation}
\int_0^{\infty} dx e^{-bx} x^{s-1} = e^{-s\ln b}\Gamma(s)~,
\label{BR1}
\end{equation}
with the condition $\textrm{Re} (b)>0$ and $\textrm{Re} (s)>0$. To satisfy these conditions for integration (\ref{BR2}), we take $b=-i(x/\hbar)+\epsilon$ and identify $s=-i(E/\tilde{\gamma})+(1/2)$ with $\epsilon>0$. At end of the integration we consider the limit $\epsilon\rightarrow 0$. This leads to
\begin{eqnarray}
\langle x|E\rangle_{f}&=&\frac{\hbar^{-\frac{iE}{\tilde{\gamma}}}}{\sqrt{2\pi}}e^{\frac{i\pi}{4}}e^{\frac{\pi E}{2\tilde{\gamma}}}\Gamma\left(\frac{1}{2}-\frac{iE}{\tilde{\gamma}}\right)\frac{x^{-\frac{1}{2}}+\frac{iE}{\tilde{\gamma}}}{\sqrt{2\pi\tilde{\gamma}}}\nonumber\\
&=&\frac{\hbar^{-\frac{iE}{\tilde{\gamma}}}}{\sqrt{2\pi}}e^{\frac{i\pi}{4}}e^{\frac{\pi E}{2\tilde{\gamma}}}\Gamma\left(\frac{1}{2}-\frac{iE}{\tilde{\gamma}}\right)\langle x|E\rangle_{i}~.
\end{eqnarray}
In the last step (\ref{energy eigenstate x rep}) has been used.
So we find the relation between the final energy eigenket $|E\rangle_{f}$ and the initial one $|E\rangle_{i}$ as
\begin{eqnarray}
|E\rangle_{f}=\underbrace{\frac{\hbar^{-\frac{iE}{\tilde{\gamma}}}}{\sqrt{2\pi}}e^{\frac{i\pi}{4}}e^{\frac{\pi E}{2\tilde{\gamma}}}\Gamma\left(\frac{1}{2}-\frac{iE}{\tilde{\gamma}}\right)}_{{C_i}}|E\rangle_{i}~.
\label{S matrix component}
\end{eqnarray}
The modulus square of the coefficient ($C_{i}$) in the above equation gives the probability of finding the particle in the initial state itself. Therefore, the transition probability for the particle to jump from initial ($|E\rangle_{i}$) to the final state ($|E\rangle_{f}$) is
\begin{eqnarray}
P = 1-|C_{i}|^{2}=\frac{1}{e^{\frac{2\pi E}{\hbar\kappa}}+1}~,
\label{transmission probability Ei to Ef}
\end{eqnarray}  
which yields again the thermal nature with temperature is given by (\ref{TH}).

Note that above expression is Fermionic in nature. This is because it is argued in \cite{Betzios:2016yaq} that $xp$ kind Hamiltonian has intrinsic Fermionic feature. We know that $xp$ can be converted to the Hamiltonian of IHO in a new pair of canonically conjugate variables. Hermitian matrix quantum model in IHO revels that the corresponding wave functions are anti-symmetric. In this section the whole analysis has been done using the energy eigenstates of the Hamiltonian. Moreover we already noticed in (\ref{xp in coordinate rep}) that the $xp$ Hamiltonian has first order derivative and so the Schrodinger like equation also has the first order derivative. Consequently corresponding Lagrangian, providing this Schrodinger like equation, bears derivative of wave function at the linear order (see discussion in \cite{Betzios:2016yaq}). All these indicate the inherent Fermionic feature of the present Hamiltonian and which may be the possible cause for the above distribution to be similar to that of Fermions.  
 
  Now, it is well known that the scattering phenomenon in black holes can provide the information about the frequency of the QNM (see page no. 397 of \cite{Paddybook}).  The imaginary part of the frequency is determined by the poles of the Gamma function appearing in Eq. (\ref{S matrix component}). It is clear that the poles are at  $E_{n}=-i\tilde{\gamma}\left(n+1/2\right)$ with $n=0,1,2\dots$. So the imaginary part of frequency is given by $\omega_n=-i\kappa(n+1/2)$ which matches with the earlier finding \cite{Kokkotas:1999bd,Berti:2009kk,Konoplya:2011qq,Paddybook}. 
  
  In this context, it is worth to mention that the probability expression (Eq. (\ref{transmission probability Ei to Ef})), obtained using the scattering process, has an intimate relationship with the probability (Eq. (\ref{Pout/Pin})) which we got using the tunneling approach in Section \ref{Tunneling formalism I}. The transition amplitude $\sim \bra{x_2}e^{-(i/\hbar)Ht}\ket{x_1}$ (known as propagator), in scattering process, is related to the Feynman's path integral $\sum_{\textrm{All paths}}\exp\Big[(i/\hbar)S\Big]$, where $S$ is the classical action. Modulus square of this quantity yields the transition probability, which is (\ref{transmission probability Ei to Ef}) in the present case. In the semi-classical limit, under saddle point approximation, path integral comes out to be proportional to $e^{(i/\hbar) S}$. Interestingly in tunneling formalism, based on WKB approximation, the ansatz for wave function is given by (\ref{phi}) which is similar to this semi-classical transition amplitude in the scattering process. Therefore it is expected that in the semi-classical regime both the tunneling probability and transition probability must coincide (For details, see Chapter $7$ of \cite{Dasbook}). It is known that this regime is best achieved by taking $\hbar\rightarrow 0$ and in this limit one can check that the probability distribution (Eq. (\ref{transmission probability Ei to Ef})), keeping only the dominating term, turns out to be  
\begin{eqnarray}
P\simeq \exp\left(-\frac{2\pi E}{\hbar\kappa}\right)~.
\end{eqnarray}
This is exactly identical to Eq. (\ref{Pout/Pin}) and thereby validating the standard relationship between scattering amplitude and the tunneling probability in the semi-classical limit. 
 

\section{\label{Perturbation}Thermality: a perturbative approach}
Here we visualise the whole system as a following effective quantum mechanical model. We first consider a free massless particle in Minkowski spacetime whose Hamiltonian is given by $H_0=p$ (with the choice of unit $c=1$). The near horizon Hamiltonian $H\simeq \kappa xp$ is treated as a small interaction of the particle with a potential of this form. So we model the actual system effectively as an interaction picture where a massless particle is interacting with the potential $\kappa xp$ when it is following the trajectory (\ref{radial 1+3D}). So we take the interaction Hamiltonian as 
\begin{equation}
\hat{H}_I=\frac{1}{2}\kappa (\hat{x}\hat{p}+\hat{p}\hat{x}) \delta(x-\frac{1}{\kappa} e^{\kappa t})~.
\label{BRM14}
\end{equation} 
Dirac-delta function has been introduced in order to make sure that the interaction is occurring only when the particle is moving along the path, given by (\ref{radial 1+3D}).  Now, if the particle is a two-level quantum atom, then there is a possibility of transition from one state to another state. Here we want to calculate the probability of transition if the atom is initially in the ground state.
So the total Hamiltonian for this quantum system is
\begin{eqnarray}
\hat{H}=\hat{H}_{0}+\hat{H}_{I}~,
\label{total Hamiltonian}
\end{eqnarray} 
where $\hat{H}_{I}$ is treated as small compared to $\hat{H}_0$. So the transition amplitude can be evaluated perturbative way. The unperturbed energy eigen basis  are evaluated from $\hat{H}_0=\hat{p}$. This will provide the initial and final basis states. These are given by $\psi_{i}(x)\sim e^{i\omega_{i}x}$ and $\psi_{f}(x)\sim e^{i\omega_{f}x}$, respectively (considering $\hbar=1$ and the velocity of light in free space $c =1$). Introducing the transition frequency $\omega=\omega_{f}-\omega_{i}$ 
we write the transition amplitude at the first order perturbation as
\begin{eqnarray}
&&c_{i\rightarrow f}=-i\int_{-\infty}^{\infty}dt\langle f|\hat{H_{I}}(t)|i\rangle ~e^{i\omega t} \delta(x-\frac{1}{\kappa} e^{\kappa t})\\
&&=-\frac{i\kappa}{2}\int_{-\infty}^{\infty}dt\langle f|(\hat{x}\hat{p}+\hat{p}\hat{x})|i\rangle e^{i\omega t}\delta(x-\frac{1}{\kappa} e^{\kappa t}).
\label{transition amp perturb}
\end{eqnarray}
Now, let us concentrate on
\begin{equation}
\mathcal{I}=\frac{\kappa}{2}\left(\underbrace{\langle f|\hat{x}\hat{p}|i\rangle\delta(x-\frac{1}{\kappa} e^{\kappa t})}_{\mathcal{I}_{1}} + \underbrace{\langle f|\hat{p}\hat{x}|i\rangle\delta(x-\frac{1}{\kappa} e^{\kappa t})}_{\mathcal{I}_{2}}\right).
\label{transition int form}
\end{equation}
The first term can be evaluated as follows:
\begin{eqnarray}
\mathcal{I}_{1}&&=\int_{-\infty}^{\infty}\langle f|x\rangle\langle x|\hat{x}\hat{p}|i\rangle \delta(x-\frac{1}{\kappa}e^{\kappa t}) dx
\nonumber
\\
&&=\int_{-\infty}^{\infty}\langle f|x\rangle x \langle x|\hat{p}|i\rangle \delta(x-\frac{1}{\kappa}e^{\kappa t})dx
\nonumber
\\
&&=\int_{-\infty}^{\infty}~\psi_{f}^{*}(x) x \left(-i\frac{\partial}{\partial x}\right)\langle x |i\rangle \delta(x-\frac{1}{\kappa}e^{\kappa t}) dx 
\nonumber
\\
&&=-i\int_{-\infty}^{\infty}~\psi_{f}^{*}(x) x \frac{\partial}{\partial x}\psi_{i}(x)\delta(x-\frac{1}{\kappa}e^{\kappa t}) dx~.
\end{eqnarray}  
In the similar approach the other term of Eq. (\ref{transition int form}) yields
\begin{eqnarray}
\mathcal{I}_{2}&=&\langle f|\hat{p}\hat{x}|i\rangle \delta(x-\frac{1}{\kappa} e^{\kappa t})= (\langle i|\hat{x}\hat{p}|f\rangle)^{*}\delta(x-\frac{1}{\kappa} e^{\kappa t})
\nonumber
\\
&=& i\int_{-\infty}^{\infty}\psi_{i}(x) x \frac{\partial}{\partial x}\psi_{f}^{*}(x) \delta(x-\frac{1}{\kappa}e^{\kappa t})dx~.
\end{eqnarray}
Then, using these and substituting the values of $\psi_{i}$ and $\psi_{f}$ along with their conjugates in (\ref{transition int form}) we obtain
\begin{eqnarray}
\mathcal{I}&&=\frac{i\kappa}{2}\int_{-\infty}^{\infty}dx\Bigg[e^{i\omega_{i}x}(-i\omega_{f})e^{-i\omega_{f}x}-e^{-i\omega_{f}x}(i\omega_{i})e^{i\omega_{i}x} \Bigg]
\nonumber
\\
&&~~~~~~~~~~\times x\delta(x-\frac{1}{\kappa}e^{\kappa t})
\nonumber
\\
&&=\frac{\kappa}{2}\int_{-\infty}^{\infty} dx e^{-i\omega x}(\omega_{f}+\omega_{i})x\delta(x-\frac{1}{\kappa}e^{\kappa t})\nonumber\\
&&=\frac{\kappa}{2}(\omega_{f}+\omega_{i})e^{-i\frac{\omega}{\kappa}e^{\kappa t}}\frac{1}{\kappa}e^{\kappa t}~.
\end{eqnarray}
\begin{widetext}
Next, using the above expression in Eq. (\ref{transition amp perturb}) and performing the integration one finds
\begin{eqnarray}
c_{i\rightarrow f}=-\frac{i(\omega_{f}+\omega_{f})}{2}\exp{\Bigg[-\Bigg(1+\frac{i\omega}{\kappa}\Bigg)\ln{\Bigg|\frac{\omega}{\kappa}\Bigg|}-\Bigg(1+\frac{i\omega}{\kappa}\Bigg)\frac{i\pi}{2}\text{sign}\Bigg(\frac{\omega}{\kappa}\Bigg) \Bigg]\Gamma\left(1+\frac{i\omega}{\kappa}\right)}~.
\label{transition prob pertrb final form}
\end{eqnarray}
In the above ``$\textrm{sign}$'' denotes the sign function.
\end{widetext}
Therefore the probability of transition from $|i\rangle$ to $|f\rangle$ turns out to be
\begin{eqnarray}
|c_{i\rightarrow f}|^{2}=\frac{\pi\kappa(\omega_{i}+\omega_{f})^{2}}{2\omega}\frac{1}{e^{\frac{2\pi\omega}{\kappa}}-1}\label{transition probability perturb}~.
\end{eqnarray}
This transition probability is thermal is nature and one identifies the temperature as (\ref{TH}).

Now let us give some physical aspects of this perturbation method. The main motive of this approach is to build a quantum mechanical model which mimics the near horizon characteristics. In this case, we have taken the potential to be $xp$ kind, which is basically the near horizon Hamiltonian. Another important point is that we have considered a definite path for the massless particle, which is basically similar to the radial trajectory of a massless particle, which we have shown already in the previous sections (Sections \ref{radial}, \ref{Covariant}, \ref{Hamiltonian Approach}). Therefore, in a physical sense, this model basically mimics the quantum behavior of the massless particle whenever it comes into the vicinity of the horizon. The non-zero value of the probability whose nature is similar to the Planckian distribution tells us about the thermal behavior in the near-horizon region. Therefore, it can be regarded as an effective approach to show the thermal nature in the near-horizon region. 

{\section{\label{Conclusion}Conclusion}}
The reason why horizon is associated with temperature has always been a fascinating question towards the physics community. On the other hand, the recent observations \cite{Hashimoto:2016dfz,Dalui:2018qqv,Dalui:2019umw} within the theoretical framework predict the possibility of induction of chaotic behaviour in a system when it is under the influence of the horizon. There is a surge of discussion in this direction.  Interestingly, both of these phenomena are characterised by a common horizon quantity, namely the surface gravity. Therefore this ``apparent interlink'' between them may help us to uncover such properties of the horizon.  In our recent work \cite{Dalui:2019esx}, we predicted that the existence of local instability, created by the horizon, maybe a possible reason for chaotic motion as well as horizon temperature.  As a continuation, in this article, we again took up this issue with great details. Here the mentioned issues in Section \ref{Intro} have been addressed. To be concrete, we focused on a recently developed conjecture, namely, the presence of local instability in the near horizon region is responsible for providing the temperature to the horizon seen by a particular set of observers. We find that this is indeed the case. For a particular set of observers, the near horizon Hamiltonian in case of an outgoing massless particle is of $xp$ kind, which is an unstable Hamiltonian, and the quantum consequences lead us to explore that thermality emerges due to this instability. This, in turn, satisfies our claim about the relationship between instability and thermality.

Now, let us discuss briefly and summarize our results what we have achieved in this paper. 
\begin{itemize}
\item{In our earlier paper \cite{Dalui:2019esx}, we found that the radial motion of an outgoing massless particle in the Painleve coordinate system is unstable in the near-horizon region. Proceeding one step further in this paper, our prime objective was to examine this whole occurrence of this feature in a more extensive way. We started our calculation by choosing another set of coordinates, which is EF coordinates, in which the motion of the massless particle along the null trajectory has been studied. Interestingly, we have found that in this coordinate also the instability still persists for the outgoing null path. Therefore, this result suggests that there are other sets of observers other than in Painleve coordinate system, which can also predict a similar instability.}
\item{Moreover, in our next investigation, using the Raychaudhuri equation, we have calculated the expansion parameter of the null geodesics, which are followed by our test particle in that particular EF coordinate. Here, we have found that the expansion parameter shows instability in the near-horizon region. It means that in the near-horizon region, instability in the particle motion is an observer independent phenomena for this particular motion of the particle.}
\item{Following the unstable nature of the horizon, next, we have constructed the Hamiltonian of the system just by the knowledge of the nature of instability in the particle motion. Interestingly, we have found that the structure of the Hamiltonian comes out to be of $xp$ kind to the observer associated with the frame of the particle where the radial motion of the particle grows as $r\sim e^{\kappa t}$ with EF time. It suggests that although the instability is observer independent for our particle motion, the particular form of the Hamiltonian is observer dependent. The observer associated with this specific frame of the particle, either in Painleve or in EF coordinates will see this form of the near horizon Hamiltonian.}
\item{After obtaining a clear picture of instability in the classical scale, we next targeted quantum calculations in order to see how thermality appears to our EF observers due to this unstable $xp$ kind structure of the near horizon Hamiltonian. We started with the tunnelling approach, where we found that this near horizon Hamiltonian predicts a finite probability of escaping the particle from the horizon and thereby providing a temperature to the horizon. The expression came out to be as that of Hawking \cite{Hawking:1974rv}.}
\item{The next approach was the detector response approach in order to get a distinct idea about the relevant vacuum state. The observer or rather the detector, in this case, is following the same null trajectory in EF coordinate in the near horizon regime, as we mentioned earlier. The vacuum was chosen to be that Boulware vacuum in this case. After evaluating the response function numerically, we have obtained that the transition probability of the detector of detecting a photon in the Boulware vacuum is similar to Planck distribution. It showed  that the detector will see  the Boulware vacuum as a thermal bath.}
\item{The other feature of the unstable potential is, it shows scattering phenomena. Therefore, our next approach was to study the scattering phenomena in the presence of this unstable $xp$ kind near horizon Hamiltonian. Identifying the ``in'' and ``out'' states we obtained the transition probability for the particle to jump from the initial to final energy state, which yielded the thermal nature again with the desired Hawking temperature. Moreover, we gained the information about the frequency of the quasinormal modes from this scattering which matches with the earlier findings \cite{Kokkotas:1999bd,Berti:2009kk,Konoplya:2011qq,Paddybook}.}
\item{In order to complete our discussion about thermality, our last approach was to consider the near horizon system as a quantum mechanical model. We built the model by considering that our near horizon $xp$ kind Hamiltonian as a small interaction Hamiltonian, and it represents the interaction with a free massless particle. We calculated the transition probability using the first-order perturbation approximation and found that the expression is thermal in nature. The motive of this approach was to construct a simple quantum mechanical model which mimics the near horizon characteristics. It turns out that this model definitely mimics the near horizon feature and can be regarded as an effective approach to show the thermal nature in the near-horizon region.}
\end{itemize}

Therefore, a clear recipe has been presented in this article about the relationship between instability and thermality in the context of horizon. However, the specific unstable spacetime region is very small, confined only within the neighbourhood of the horizon. Moreover, it is also shown that in the quantum regime, this instability provides the automatic emergence of the temperature to the system, which is exactly equal to the Hawking expression. The way we have studied here is very interesting. Within the various known techniques, the black hole system has been investigated. On this note, we feel that the results, as well as the techniques, introduced here, will not only have a significant impact in the area of black hole physics, but also may uncover several unknown sides of the horizon. Furthermore, the present discussion has been confined within a static, spherically symmetric black hole. So it would be interesting if the same can be extended to Kerr and other non-trivial backgrounds. The investigations in these directions are in progress and hope we will be able to report soon.

\noindent
{\bf Acknowledgements}
\vskip 1mm
\noindent
{\scshape{We dedicate this work to all the warriors against COVID-19 across the globe.}}
\vskip 3mm

\appendix
\section{{\label{App1}}Evaluation of $\tilde{\kappa}$, $\Theta$ and $\sigma_{ab}$ for the null vector (\ref{lalpha}) in the background (\ref{E-F metric})}
{\it Non-affinity coefficient $\tilde\kappa$} -- 
Consider the null normal vector field $l_{a}$ of any null hypersurface generates a null geodesic congruence. The non-affinely parametrised geodesic equation is given by
\begin{eqnarray}
l^{b}\nabla_{b}l_{a}=\tilde{\kappa}l_{a}~,
\label{BRM11}
\end{eqnarray}
where $\tilde{\kappa}$ is called the non-affinity coefficient. In order to find $\tilde{\kappa}$ for the geodesic curves, given by (\ref{lalpha}), first we need to compute the gradient of the null normal ($\nabla_{b}l_{a}$). We have already computed the components of $l_{a}$ in Section (\ref{Null hypersurface}) (see Eq. (\ref{l_alpha})).  Therefore, the components of $\nabla_{b}l_{a}$ in EF coordinates are obtained for the metric (\ref{E-F metric}) as
\begin{widetext}
\begin{eqnarray}
\nabla_{b}l_{a}=
\begin{pmatrix}
\frac{1}{2}\frac{f(r)f'(r)}{f(r)-2} & \frac{1}{2}\frac{(f(r)-4)f(r)f'(r)}{(f(r)-2)^{2}} & 0 & 0 \\
\frac{f'(r)}{2} & \frac{1}{2}\frac{(f(r)-4)f'(r)}{f(r)-2} & 0 & 0\\
0 & 0 & -\frac{1}{2}\frac{f(r)h'(r)}{f(r)-2} & 0 \\
0 & 0 & 0 & -\frac{1}{2}\frac{f(r)h'(r)}{f(r)-2}\sin^{2}\theta
\end{pmatrix}~.
\label{gradient_of_l_components}
\end{eqnarray}
\end{widetext}
Using the values of the $\nabla_{b}l_{a}$ components we obtain
\begin{eqnarray}
l^{b}\nabla_{b}l_{a}=\left(\frac{2f'(r)f(r)}{(f(r)-2)^{3}}, \frac{2f'(r)}{(f(r)-2)^{2}},0,0\right)~.
\label{geodesic_eqn}
\end{eqnarray}
Using the geodesic equation (\ref{BRM11}) and comparing Eq. (\ref{geodesic_eqn}) with the expression of Eq. (\ref{l_alpha}), we deduce the value of $\tilde{\kappa}$ as
\begin{eqnarray}
\tilde{\kappa}=\frac{2f'(r)}{(f(r)-2)^{2}}~.
\label{kappa}
\end{eqnarray} 

{\it Expansion parameter $\Theta$} --
The expansion parameter of the congruence of geodesics $\Theta$ is defined as
\begin{eqnarray}
\Theta = q^{ab}\nabla_al_b~,
\label{define theta}
\end{eqnarray}
where $q_{ab}$ is the transverse part of the metric $g_{ab}$, defined as
\begin{eqnarray}
q_{ab}=g_{ab}+l_{a}n_{b}+l_{b}n_{a}~.
\label{induced metric}
\end{eqnarray}
Here $n_{a}$ is an auxiliary null vector field which satisfies $l^{a}n_{a}=-1$. Therefore using (\ref{BRM11}) and (\ref{induced metric}) in (\ref{define theta}), the parameter $\Theta$ can be expressed in terms of $l^{a}$ and $\tilde{\kappa}$ as
\begin{eqnarray}
\Theta &=&\nabla_{a}l^{a}-\tilde{\kappa}
\nonumber
\\
&=&\frac{1}{\sqrt{-g}}\partial_{a}(\sqrt{-g}~l^{a}) - \tilde{\kappa}~,
 \label{theta}
\end{eqnarray} 
where $g$ is the determinant of the metric. Since, in this case the determinant of the metric (\ref{E-F metric}) is $g=-h^{2}(r)\sin^{2}\theta$ and $l^{a}$ is given by (\ref{lalpha}), Eq.(\ref{theta}) yields 
\begin{eqnarray}
\Theta &=& \frac{1}{h(r)\sin\theta}\partial_{r}\left[\frac{h(r)f(r)}{2-f(r)}\sin\theta\right]-\tilde{\kappa}~.
\label{theta1}
\end{eqnarray}
Next using (\ref{kappa}) in the above we obtain the expression of the expansion parameter
\begin{eqnarray}
\Theta=\frac{h'(r)f(r)}{h(r)(2-f(r))}~.
\label{theta expression}
\end{eqnarray}

{\it Shear parameter $\sigma_{ab}$} --
The shear parameter $\sigma_{ab}$ of the congruence of geodesics is defined as
\begin{eqnarray}
\sigma_{ab}=\frac{1}{2}\left(b_{ab} + b_{b a}-\Theta q_{ab}\right)~,
\label{Sigma}
\end{eqnarray}
where $b_{ab}$ is the orthogonal component of $\nabla_al_b$ projected by $q_{ab}$:
\begin{eqnarray}
b_{ab}=q_{a}^{c}q_{b}^{d}\nabla_cl_d~.
\label{transverse_comp_B}
\end{eqnarray}
Using (\ref{gradient_of_l_components}) in (\ref{transverse_comp_B}) one can easily calculate each component of $\sigma_{ab}$. This can be readily shown that each term of $\sigma_{ab}$ vanishes i.e.
\begin{eqnarray}
\sigma_{ab}=0~.
\label{sigma}
\end{eqnarray}

\section{{\label{App3}}Detector's response in $(1+1)$ dimensional Schwarzschild background}
In section \ref{Near horizon detector response} we studied the transition probability for an atomic detector, which is interacting with a massless scalar field, moving very close to the horizon. It is found that it will register a particle in the Boulware vacuum. This was done numerically. Here we shall present an analytical approach when the detector is moving throughout the spacetime along our chosen null path, which near to the horizon leads to an unstable trajectory (\ref{radial 1+3D}). The metric will be chosen to be Schwarzschild black hole in $(1+1)$ spacetime dimensions. The two-dimensional case is analytically solvable, and since we will be interested in finding the detected temperature, it is sufficient to consider a two-dimensional situation. Here both ingoing and outgoing detectors will be studied. We shall adopt the previous atomic detector model, and so the working formula for transition probability is given by (\ref{probability of detection in radial form}).

The Schwarzschild metric in $(1+1)$ dimensional spacetime in Schwarzschild coordinates $(t_{s},r)$ is given by
\begin{eqnarray}
ds^{2}=-f(r)dt^{2}_{s}+\frac{dr^{2}}{f(r)}~,
\label{BRM13}
\end{eqnarray}
where $f(r)=\left(1-\frac{r_{H}}{r}\right)$. The horizon located at $r_{H}=2M$ where $M$ is the mass of the black hole. In the Eddington-Finkelstein coordinates $(t,r)$ the metric transforms into
\begin{equation}
ds^{2}=-\left(1-\frac{r_{H}}{r}\right)dt^{2}+\frac{2r_{H}}{r}dtdr+\left(1+\frac{r_{H}}{r}\right)dr^{2}~.
\label{Schwarzschild metric}
\end{equation}  
In this case the tortoise coordinate is given by
\begin{equation}
r_* = r+r_H\ln\Big(\frac{r}{r_H}-1\Big)~.
\label{BRM12}
\end{equation}
In the following calculation we shall choose the unit such that $\hbar=1$.

\subsection{\label{Detector response in}Outgoing detector}
The outgoing null path can be determined as earlier. The detector is moving from horizon to radial infinity. In this case the tangent of the path is 
is determined by the $t$ and $r$ components of (\ref{lalpha}). Therefore the path is found to be the solution of
\begin{eqnarray}
\frac{dr}{dt}=\frac{\frac{r}{r_{H}}-1}{\frac{r}{r_{H}}+1}~.
\label{radial eqn for schild}
\end{eqnarray} 
Performing the above integration we obtain
\begin{eqnarray}
t=r+2r_{H}\ln{\left[\frac{r}{r_{H}}-1\right]}~.
\label{t for schild}
\end{eqnarray}
We have already found the positive frequency mode corresponding to the detector (see Eq. (\ref{detector mode})). Next, we need to find the positive frequency Boulware mode for the massless scalar field i.e., $u_{\nu}$, and for that, we need to solve the Klein-Gordon (KG) equation $\square\phi=0$ under the background (\ref{Schwarzschild metric}). Since the detector is outgoing, the scalar mode under investigation will be ingoing one.
This is given by (\ref{mode solution for photon}).
Now, substituting everything in the general form (\ref{probability of detection in radial form}) (i.e.  use (\ref{t for schild}) and (\ref{mode solution for photon}) along with (\ref{BRM12})) with $\tau=t$ and re-expressing it in terms of the radial coordinate we obtain   
\begin{eqnarray}
P_{\uparrow}=Q^{2}\Big|\int_{r_{H}}^{\infty}dr&&\left(\frac{\frac{r}{r_{H}}+1}{\frac{r}{r_{H}}-1}\right)e^{i(2\nu+\omega)r}\nonumber\\&&\times\left(\frac{r}{r_{H}}-1\right)^{2i r_{H}(\nu+\omega)} \Big|^{2}~.
\label{response 1}
\end{eqnarray}
Changing the variable as $(\frac{r}{r_{H}}-1)=y$, we find
\begin{eqnarray}
P_{\uparrow}&=&Q^{2}~\Big|r_{H}\int_{0}^{\infty}dy~\left(\frac{y+2}{y}\right)~e^{i(2\nu+\omega)r_{H}(y+1)}\nonumber\\
&&\times~ y^{2ir_{H}(\nu+\omega)}\Big|^{2}
\nonumber
\\
&=&Q^{2}|I_{\uparrow 1} + I_{\uparrow 2}|^{2}~,
\label{response 2}
\end{eqnarray}
where
\begin{eqnarray}
I_{\uparrow 1}= r_{H}e^{i(2\nu+\omega)r_{H}}\int_{0}^{\infty}dy~ y^{2ir_{H}(\nu+\omega)}\nonumber\\
\times e^{i(2\nu+\omega)r_{H}y}
\label{I1}
\end{eqnarray}
and
\begin{eqnarray}
I_{\uparrow 2}= 2 r_{H}e^{i(2\nu+\omega)r_{H}}\int_{0}^{\infty}dy~ y^{2ir_{H}(\nu+\omega)-1}\nonumber\\
\times e^{i(2\nu+\omega)r_{H}y}~.
\label{I2}
\end{eqnarray}
These integrations can be performed using the general formula (\ref{BR1}) and following the prescription, as performed in section \ref{BRQNM}.
This leads to
\begin{widetext}
\begin{equation}
I_{\uparrow 1}=r_{H}e^{i(2\nu + \omega)r_{H}}\exp{ \Bigg[-(1+2ir_{H}(\nu+\omega))\Bigg(\ln{|(2\nu+\omega)r_{H}|}-\frac{i\pi}{2}\text{sign}[(2\nu+\omega)r_{H}]\Bigg)\Bigg]}\Gamma\left(1+2ir_{H}(\nu+\omega)\right)
\label{I1 value}
\end{equation}
and
\begin{equation}
I_{\uparrow 2}=2 r_{H}e^{i(2\nu + \omega)r_{H}}\exp{ \Bigg[-2ir_{H}(\nu+\omega)\Bigg(\ln{|(2\nu+\omega)r_{H}|}-\frac{i\pi}{2}\text{sign}[(2\nu+\omega)r_{H}]\Bigg)\Bigg]}\Gamma\left(2ir_{H}(\nu+\omega)\right)~.
\label{I2 value}
\end{equation}
\end{widetext}
Substituting them in Eq. (\ref{response 2}) and performing the modulus square, we finally obtain the expression for the transition probability as 
\begin{eqnarray}
P_{\uparrow}=Q^{2}\frac{4\pi r_{H}\nu^{2}}{(2\nu+\omega)^{2}(\nu+\omega)}\times\frac{1}{e^{4\pi r_{H}(\nu+\omega)}-1}~.\label{response final}
\end{eqnarray}
This is thermal in nature and the temperature is identified as 
\begin{eqnarray}
T=\frac{1}{4\pi r_{H}}~,
\label{temperature schild}
\end{eqnarray}
which is the Hawking expression for Schwarzschild black hole.

\subsection{\label{Detector response out}Ingoing detector}
The detector is now approaching towards the horizon from radial infinity. In this case, the null trajectory is chosen to be along the tangent, which is normal to ingoing null Krushkal-Szekeres coordinate $V=$ constant surface. This is defined by
\begin{eqnarray}
V=\pm \exp(\kappa v)+1~,
\end{eqnarray}
where $V=1$ is the horizon.
The observer's coordinates are chosen to be outgoing Eddington-Finkelstein coordinates $(u,r)$. Then the  EF timelike coordinate $(t)$ is given by Eq. (\ref{E-F time outgoing}). In these coordinates $(t,r)$ the metric (\ref{BRM13}) takes the following form:
\begin{equation}
ds^{2}=-f(r)dt^{2}+2(f(r)-1)dt\,dr+(2-f(r))dr^{2}~.
\label{E-F metric outgoing}
\end{equation}
Now as earlier, the tangent to the path is given by
\begin{eqnarray}
l^{a}=\left(1,\frac{f(r)}{f(r)-2}\right)~.
\label{lalpha out}
\end{eqnarray}
Correspondingly the covariant components are
\begin{eqnarray}
l_{a}=\left(\frac{f(r)}{f(r)-2},-1\right)~.
\label{l_alpha out}
\end{eqnarray}
Therefore again the detector is moving along the radial direction only and the trajectory is determined by 
\begin{eqnarray}
\frac{dr}{dt}&=&\frac{1-\frac{r}{r_{H}}}{1+\frac{r}{r_{H}}}~.
\label{dr/dt out}
\end{eqnarray}
Performing the integration in Eq. (\ref{dr/dt out}) we obtain the solution of $t$ as
\begin{eqnarray}
t=-r-2r_{H}\ln{\left[\frac{r}{r_{H}}-1\right]}~.\label{t out}
\end{eqnarray}
Since the detector is ingoing, we shall investigate the outgoing Boulware scalar mode, given by 
\begin{eqnarray}
u_{\nu}=e^{-i\nu(t_{s}-r_{*})}~.
\end{eqnarray} 
Substituting all these in (\ref{probability of detection in radial form}) and proceeding in the previous way one finds that the transition probability is same as (\ref{response 1}). Therefore the final expression is given by (\ref{response final}). Hence the ingoing detector will register particle in the Boulware vacuum with Hawking temperature (\ref{temperature schild}). 
\section{\label{Through Gutzwiller}A note on thermality through Gutzwiller's formula} 
In our earlier work \cite{Dalui:2019esx} the quantum thermal character of our $xp$ Hamiltonian was revealed using Gutzwiller's formula \cite{Stockmann:1999,Gutzwiller:1990}
\begin{equation}
\mathbf{g}(E)=-\frac{i}{\hbar}\sum_{l}\frac{T_{l}}{\vert\vert M_{BA,l} -1\vert\vert^{\frac{1}{2}}}
\exp\bigg[\frac{i}{\hbar}S_{l}(E)-i\frac{\mu_{l}\pi}{2}\bigg]~,
\label{Green_function}
\end{equation}
to derive the density of states (DOS) of a system. The DOS is given by 
\begin{eqnarray}
\rho(E)=-\frac{1}{\pi}\text{Im}(\mathbf{g}(E))~.
\label{density of states}
\end{eqnarray}
The meaning of each of the term is given in \cite{Dalui:2019esx}. We already showed in \cite{Dalui:2019esx} that for our Hamiltonian the DOS is thermal in nature. The whole focus was to evaluate the Jacobi action
\begin{equation}
S_l(E) = \oint pdx~,
\label{BRM15}
\end{equation}
for the $l^{th}$ closed orbit. Since the Hamiltonian corresponds to an unstable trajectory, in order to perform the closed integration in (\ref{BRM15}) an analytic continuation to complex plane approach was adopted.  Transforming to IHO Hamiltonian, the frequency like quantity was complexified which led to harmonic oscillator Hamiltonian. The trajectory is now closed and the action was obtained to be as
\begin{eqnarray}
S_{l}(E)=\frac{2\pi i E_l}{\kappa}~.
\label{action 1}
\end{eqnarray}
This yielded the thermal density of states as
\begin{eqnarray}
\rho(E)=\frac{1}{\hbar \kappa}\sum_{l}\frac{1}{\sinh{\pi}} e^{-\frac{2\pi E_{l}}{\hbar\kappa}}\cos{\frac{\mu_{l}\pi}{2}}~,
\label{density of states for xp}
\end{eqnarray}
with the temperature is given by (\ref{TH}).
See \cite{Dalui:2019esx} for detailed calculation.

Here we shall show that the action (\ref{action 1}) can also be obtained in a different way. Since our Hamiltonian (\ref{xp Hamiltonian}) is valid very near to the horizon, we consider a closed path which encircles the horizon $x=0$ in a circular trajectory with a very small radius (say $\epsilon\rightarrow 0$) as shown in Fig.  \ref{Fig3}.

\begin{figure}[h]
\centering
\includegraphics[scale=0.2]{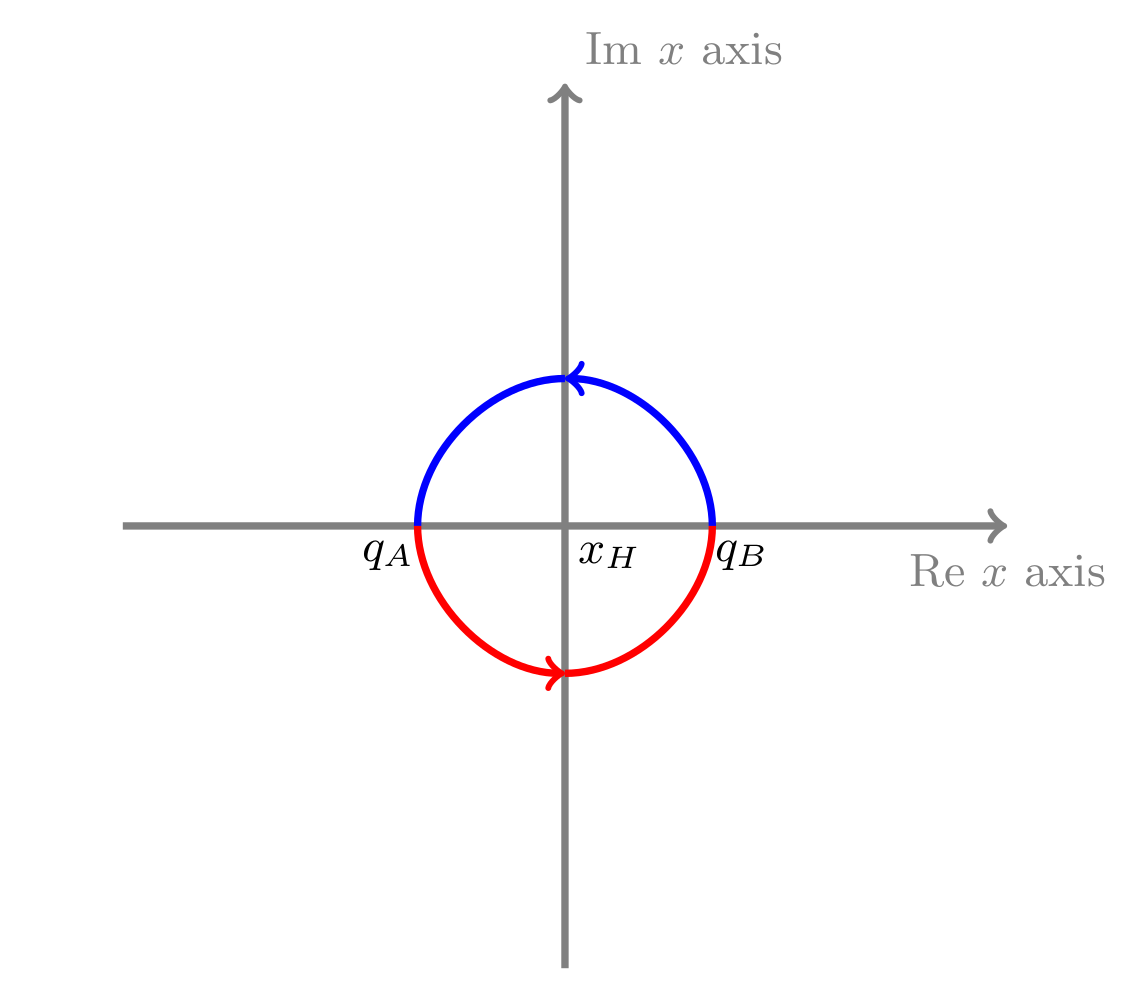} 
\caption{The contour diagram across the horizon where the horizon is at $x_{H}$}\label{Fig3}
\end{figure}
\noindent
Actually the path is one which starts just outside the horizon, enters through it and comes back again. So it crosses the singular point $x=0$ twice. To avoid this a complex path has been chosen and since the relevant contribution comes from the singularity, we have chosen a circular path as shown in Fig. \ref{Fig3}.
The choice of these types of paths motivated from the semi-classical treatment of Hawking effect in tunneling formalism, similar to what we already discussed in Section \ref{Tunneling formalism I}. Since the formula (\ref{Green_function}) is semi-classical in nature, we hope that such paths are relevant here as well.
With this the action of the particle following the closed path is  
\begin{eqnarray}
S_l(E)=\oint p~dx~=\frac{E_l}{\kappa}\oint\frac{dx}{x}~.
\label{action for closed path}
\end{eqnarray}
The above closed integration can be divided into two parts: 
\begin{eqnarray}
S_l(E)&=&\frac{E_l}{\kappa}\Bigg[\underbrace{\int_{q_{A}}^{q_{B}}\frac{dx}{x}}_{I_{1}}+\underbrace{\int_{q_{B}}^{q_{A}}\frac{dx}{x}}_{I_{2}}\Bigg]\label{action}
\end{eqnarray}
Now, the integration $I_{1}$ i.e. when the particle is going from $q_A$ to $q_B$ is evaluated as 
\begin{eqnarray}
I_{1}=\int_{q_{A}\rightarrow q_{B}}\frac{dx}{x}
=\int_{\pi}^{2\pi}\frac{i\epsilon e^{i\vartheta}}{\epsilon e^{i\vartheta}}d\vartheta 
=i\pi~,
\end{eqnarray}
where in the above $\epsilon$ is chosen to be the radius of the circular path and we substituted $x=\epsilon e^{i\vartheta}$.
Similarly, $I_{2}$ is evaluated to be
\begin{eqnarray}
I_{2}=i\pi~.
\end{eqnarray}
Finally, putting the values of $I_{1}$ and $I_{2}$ in (\ref{action}), we obtain the same expression of the action as in (\ref{action 1}). 


\end{document}